\documentclass[5p]{elsarticle}
\usepackage{cancel}
\usepackage{makeidx}
\usepackage{amsthm}
\usepackage{amssymb}
\usepackage{amsfonts}
\usepackage{graphicx}
\usepackage{slashed}
\usepackage{amsmath}
\usepackage{multirow}
\usepackage{array}

\begin{document}

\begin{frontmatter}

\title{Bott Periodicity and Realizations of Chiral Symmetry in Arbitrary Dimensions}

\author[]{Richard DeJonghe}

\author[]{Kimberly Frey}

\author[]{Tom Imbo\corref{cor}}
\ead{imbo@uic.edu}

\cortext[cor]{Corresponding author}
\address{Department of Physics, University of Illinois at Chicago, 845 W. Taylor St., Chicago, IL 60607

\vspace{-.3in}
}

\begin{abstract}
We compute the chiral symmetries of the Lagrangian for confining ``vector-like'' gauge theories with massless fermions in $d$-dimensional Minkowski space and, under a few reasonable assumptions, determine the form of the quadratic fermion condensates which arise through spontaneous breaking of these symmetries.  We find that for each type (complex, real, or pseudoreal) of representation of the gauge group carried by the fermions, the chiral symmetries of the Lagrangian, as well as the residual symmetries after dynamical breaking, exactly follow the pattern of Bott periodicity as the dimension changes. The consequences of this for the topological features of the low-energy effective theory are considered.
\end{abstract}


\end{frontmatter}

\section{Introduction}

Spontaneous breaking of chiral symmetry is a central non-perturbative feature of QCD.  It not only explains the large effective mass of quarks bound within hadrons, but also allows one to understand pions as Goldstone bosons of this broken symmetry.  Given the importance of chiral symmetry breaking in QCD, we will investigate this phenomenon in a wider class of theories --- namely, confining vector-like gauge theories in Minkowski space of arbitrary dimension $d$ --- in an attempt to obtain a broader perspective on the nature of chiral symmetry breaking.  (What we mean by a ``vector-like'' theory for $d$ odd will be made clear below.)  For $d > 4$ certain aspects of these models\footnote{Although gauge theories in $d>4$ are not perturbatively renormalizable, we may consider such models as effective theories arising from an appropriate UV completion. However, since these gauge theories are naively free in the infrared, we assume the presence of additional degrees of freedom (or other dynamical modifications) which conspire to render the theory confining, but which do not otherwise play a role in 
determining the patterns of chiral symmetry breaking. More robustly, similar chiral symmetry breaking patterns will be obtained for any strongly coupled vector-like field theory of massless fermions (in any dimension) transforming irreducibly under an internal symmetry group~$G$, and for which an appropriate \mbox{$G$-invariant} condensate forms.  Using gauge symmetries simply makes our analysis more concrete.} may be relevant to higher dimensional extensions of the Standard Model, while some models with $d < 4$ might be relevant in condensed matter physics.

Various results are known for $d \leq 4$.  In $3+1$ dimensions, Peskin \cite{peskin_1980,peskin_slac} and Preskill \cite{preskill} have worked out the patterns of spontaneous chiral symmetry breaking (for an arbitrary gauge group $G$ and an arbitrary representation of $G$ carried by the fermions) under certain assumptions.  Different patterns of ``chiral'' symmetry breaking have been found in $2+1$ dimensions for fermions in a complex representation of any gauge group \cite{pisarski_1984,appelquist_1986,appelquist_1995,appelquist_1990}, in the fundamental (pseudoreal) representation in an $SU(2)$ gauge theory \cite{magnea_00a,hilmoine_00,nagao_01}, and in the adjoint (real) representation in an $SU(N_c)$ gauge theory \cite{hilmoine_00,nagao_01,magnea_00b}.

Our analysis  is consistent with the aforementioned results and goes well beyond them.  Under reasonable assumptions, we not only determine the form of the relevant condensates generated by dynamical breaking for arbitrary $d$ and $G$, but find that for each type (complex, real, or pseudoreal) of representation of $G$ carried by the fermions, the chiral symmetries of the massless Lagrangian, as well as the residual symmetries after dynamical breaking, exactly follow the pattern of Bott periodicity as the dimension changes. The consequences of this for the topological features of the low-energy effective theory of the Goldstone boson degrees of freedom are considered, including an analysis of the interpretation of baryons as topological solitons.

\section{Chiral Symmetries of the Lagrangian}
\label{kineticterm}

We consider a confining gauge theory in $d$-dimensional Minkowski space with compact gauge group $G$,  where the gauge fields are coupled to $N$ flavors of massless fermions which all transform under a single irreducible unitary representation $r$ of $G$.  We denote the fermion fields by $\psi^{\,i,a}$, where $i = 1, \ldots, N$ is the flavor index and $a = 1, \ldots, \dim r$ is~the ``color'' index.  The spinor index is suppressed.

When $d$ is even, we take each $\psi^{\,i,a}$ to be a Dirac spinor.  Here $\psi^{\,i,a}$ decomposes uniquely into left-handed ($\psi^{\,i,a}_L$) and right-handed ($\psi^{\,i,a}_R$) Weyl spinors, corresponding to the two inequivalent irreducible representations of the group $Spin(1,d-1)$.  When $d$ is odd, however, there is only a single irreducible representation  (up to equivalence) of $Spin(1,d-1)$; i.e.~a single type of Weyl spinor.  Hence, in order for there to be ``left-handed'' and ``right-handed'' spinors (and therefore a notion of chirality), we must consider each $\psi^{\,i,a}$ to consist of \emph{two} copies of this unique Weyl spinor, where the parity transformation is defined such that the ``left-handed'' and ``right-handed'' spinors are interchanged \cite{pisarski_1984,appelquist_1986,shimizu_1985}.  As~a~consequence, when $d$ is odd we may alternatively think of the theory as having $2N$ identical  ``Weyl flavors''.  In what follows, we will suppress flavor and color indices, and simply denote the fermion fields by~$\psi$.

A gauge theory (in any dimension) will be called \emph{vector-like} if it has the fermion content just described, and a Lagrangian which treats ``left-handed'' and ``right-handed'' spinors democratically.  In particular, we take the Lagrangian density to be
\begin{equation}
\label{eq:lagrangian}
\mathcal{L} = \bar{\psi} \slashed{D} \psi + \textrm{Tr} \ F^{\mu \nu} F_{\mu \nu} ,
\end{equation}  
where $\slashed{D} = i \gamma^{\mu} ( \partial_{\mu} + i g t^c_r A^c_{\mu} )$, with $g$ the coupling constant, $\gamma^\mu$ the appropriate Dirac matrices,\footnote{Without loss of generality, we take each of our spinor representations to arise from a representation of the $d$-dimensional Clifford algebra in which $\gamma^{0}$ is Hermitian and $\gamma^{a}$ is anti-Hermitian for all ${a \in \{ 1,2, ...d-1 \}}$.} $\bar{\psi} = \psi^\dag \gamma^0$, and $t^c_r$ the generators of the Lie algebra of $G$ in the representation associated with $r$.  We~also take the fermion fields to be Grassmann-valued.

We define the chiral symmetries to be the global symmetries of $\mathcal{L}$ which can be represented as real linear transformations\footnote{A map $Z$ on a complex vector space $V$ is called a \emph{real linear transformation} if $Z(\alpha v + \beta w) = \alpha Z(v) + \beta Z(w)$ for all $\alpha, \beta \in \mathbb{R}$ and $v,w \in V$.} $Z$ acting on the indices of $\psi$ such that $Z$ commutes with all gauge and spinor transformations.\footnote{Our analysis will not address any anomalous breaking of these symmetries, although we will briefly return to this issue at the end of section 4.}   This definition reproduces the standard notion of chiral symmetry for vector-like theories in $3+1$ dimensions \cite{peskin_1980}.  The reason we consider \emph{real} linear transformations (and not just the linear ones) is that these are more natural when $\psi$ is not in a complex representation of ${Spin(1,d-1) \times G}$.  For example, when $\psi$ carries a representation $\rho$ in which the group elements are all represented by real matrices, the real and imaginary parts of $\psi$ do not mix under $\rho$; hence, these can be thought of as independent fields.  However, the natural transformations on the carrier space of $\rho$ which allow the real and imaginary parts of $\psi$ to transform independently are not linear, but merely real linear.  For~such a $\rho$, one could alternatively compute the chiral symmetries by restricting the fermion content so that $\psi$ is real \cite{magnea_00b};  however this trick does not work when the representation is pseudoreal.  By considering real linear transformations (and complex $\psi$), we are able to treat the real, pseudoreal, and complex cases all on the same~footing.\footnote{Note that our definition of chiral symmetry can be applied to a theory with any fermion content, even one without a notion of chirality such as a confining gauge theory with $d$ odd and an odd number of Weyl flavors.  However, in this particular example we would find that (under reasonable assumptions) the dynamically generated fermion condensates would not break these flavor symmetries.  This is why, for $d$ odd, we have taken the theory to be vector-like in the above sense.  For $d$ even, on the other hand, results similar to those below can be obtained for many theories which are not vector-like.}

The chiral symmetries of $\mathcal{L}$ will depend on the representations of $G$ and $Spin(1,d-1)$ under which $\psi$ transforms.  There is a well-known classification of irreducible representations which will be useful here.  A  representation $\rho_0$ of a group $G_0$ is called \emph{real} (respectively, \emph{pseudoreal}) if there exists an antilinear operator $J$ on the carrier space of $\rho_0$ which is equivariant with respect to $\rho_0$ (i.e.~$[J,\rho_0(g)] = 0$ for all $g \in G_0$) and also satisfies $J^2 = I$ (respectively, ${J^2 = - I}$), where $I$ is the identity operator.  (Such a $J$ is unique up to a phase.)  If $\rho_0$ is neither real nor pseudoreal it is called \emph{complex}.  If $\rho_0$ is irreducible, then it falls into exactly one of these three categories.  Given a representation $\rho_1$ of some other group $G_1$, we can form the outer tensor product representation $\rho_0 \otimes \rho_1$ of $G_0 \times G_1$.  Now assume $\rho_0$ and $\rho_1$ (and hence $\rho_0 \otimes \rho_1$) are irreducible. Then, if either $\rho_0$ or $\rho_1$ is complex, $\rho_0 \otimes \rho_1$ will be as well.  If, instead, $\rho_0$ and $\rho_1$ are either both real or both pseudoreal, $\rho_0 \otimes \rho_1$ will be real.  In all other cases, $\rho_0 \otimes \rho_1$ will be pseudoreal.  For a more detailed discussion see \cite{grimus_1997}.

\subsection{Symmetries of $\mathcal{L}$ for $d$ odd}

For $d$ odd, we have that the  irreducible representation of 
${Spin(1,d-1)}$ (which we denote by $s$) is real for $d = 1,3$ (mod $8$) and pseudoreal for $d = 5,7$ (mod $8$).  In the cases in which $r$ (i.e.~the irreducible representation of the gauge group $G$) is also real or pseudoreal, the product representation ${\rho \equiv s \otimes r}$ of ${Spin(1,d-1) \times G}$ is either real or pseudoreal.  As such, there exists the antilinear equivariant map $J$ on the carrier space of $\rho$ described above.  Additionally, Schur's lemma guarantees that the scalars are the only linear maps which commute with $\rho(g)$ for all ${g \in Spin(1,d-1) \times G}$.  By extending $\rho$ to include the trivial representation on ``flavor'' space,\footnote{Recall that when $d$ is odd, we may consider our theory as having a ``Weyl flavor'' index running from $1$ to $2N$.  In what follows (for $d$ odd), ``flavor'' space will refer to the $2N$-dimensional complex vector space $\mathbb{C}^{N} \otimes \mathbb{C}^2$ associated with this Weyl index.  Note that parity operates on ``flavor'' space as $I \otimes \sigma_1$.} we have that the most general real linear transformation commuting with $\rho$ is an operator of the form $X + Y J$, where $J$ has been extended to act on ``flavor'' space simply as complex conjugation\footnote{Extending $J$ (which acts on the carrier space of $s \otimes r$) results in an operator of the form $J \otimes K$, where $K$ acts on the ``flavor'' space ${\mathbb{C}^N \otimes \mathbb{C}^2}$.  While $K = I$ may seem the natural choice, the tensor product of an antilinear operator with a linear operator is not well-defined \cite{grimus_1997}.  We~choose $K$ to be complex conjugation since it is the simplest choice for which $J \otimes K$ is the appropriate equivariant antilinear map associated with the extended $\rho$; however, any antilinear $K$ satisfying $K^2 = I$ would give equivalent results.}, and $X,Y$ are (standard) linear operators which act non-trivially only on ``flavor'' space.

One can show that $\mathcal{L}$ remains invariant under the transformation $\psi \rightarrow (X + Y J) \psi \,$  if and only if
\begin{equation}
\label{oddeq1a}
 X^{\dagger} \gamma^{0}\slashed{D} X - (Y J)^{\dagger} \gamma^{0}\slashed{D} (Y J) = \gamma^{0}\slashed{D} 
\end{equation}
and
\begin{equation}
\label{oddeq1b}
X^{\dagger} \gamma^{0}\slashed{D} (Y J) K = (X^{\dagger} \gamma^{0}\slashed{D} (Y J) K)^{T},
\end{equation}
where $K$ denotes the complex conjugation operator on the carrier space of $\rho$ (in some basis).  The conditions (\ref{oddeq1a}) and (\ref{oddeq1b}) above are equivalent to
\begin{equation}
\label{oddeq2} 
X^{\dagger} X + Y^{T} Y^{*} = I  \  \ \quad \textrm{and} \quad \ \  X^{\dagger} Y = \mp Y^{T} X^{*},
\end{equation}
where the signs correspond to $J^{2} = \pm I$. The expressions in (\ref{oddeq2}) can be obtained by using that $J$ anticommutes with $\gamma^{0}\slashed{D}$ (which follows from the fact that $J$ is equivariant with respect to $\rho$) and\footnote{If $\psi$ were c-number (rather than Grassmann) valued, we would have that $J^{\dagger} = \pm J$.} $J^{\dagger} = \mp J$, where the signs are associated with the cases $J^2 = \pm I$.  When $\rho$ is real (respectively, pseudoreal), it follows directly from (\ref{oddeq2}) that the chiral symmetry transformations form a group isomorphic to $O(4N)$ ($Sp(2N)$, respectively).\footnote{We use the convention that $Sp(n)$ is the subgroup of $U(2n)$ which commutes with $\left( \begin{array}{cc} 0 & -I \\ I & 0 \end{array} \right)$, where $I$ is the $n \times n$ identity matrix.}

In the case in which $r$ is complex, $\rho$ is also complex.  Here, no equivariant antilinear map exists, so any chiral symmetry transformation must be a (standard) linear operator $X$ which acts non-trivially only on ``flavor'' space.  It can easily be seen that the transformation $\psi \rightarrow X \psi$ preserves $\mathcal{L}$ if and only if $X$ is unitary.  Thus, when the overall representation is complex, the chiral symmetry group is given by $U(2N)$. 

\subsection{Symmetries of $\mathcal{L}$ for $d$ even}

We denote the left-handed (right-handed) spinor representation by $s_L$ ($s_R$).  For $d = 2,6$ (mod $8$), $s_L$ and $s_R$ are either both real or both pseudoreal.  In such cases, $\psi_L$ and $\psi_R$ transform independently under chiral symmetry transformations, regardless of whether $r$ is real, pseudoreal or complex.  When $r$ is either real or pseudoreal, the product representations $\rho_L \equiv s_L \otimes r$ and $\rho_R \equiv s_R \otimes r$ are either both real or both pseudoreal, so that there exists two equivariant antilinear maps $J_{L}$ and $J_{R}$ acting on the carrier spaces of $\rho_L$ and $\rho_R$, respectively.  Both $J_L$ and $J_R$ square to $\pm I$ and are unique up to a phase; however, no such maps exist when $r$ is complex. In any case, $\psi_L$ ($\psi_R$) transforms according to the representation $\rho_L$ ($\rho_R$), hereafter extended to act trivially on flavor space.  As such, when $r$ is either real or pseudoreal, any chiral symmetry transformation must be of the form $(X_L + Y_L J_{L}) \oplus (X_R + Y_R J_{R})$, where $J_{L}$ ($J_{R}$) is extended to act as complex conjugation on flavor space, and where $X_L$ and $Y_L$ ($X_R$ and $Y_R$) are (standard) linear operators which act on $\psi_L$ ($\psi_R$) and are non-trivial only on flavor space.  When $r$ is complex, the chiral symmetry transformations can only be of the form $X_L \oplus X_R$.  

Requiring $\mathcal{L}$ to be invariant under the transformations above, we obtain constraints similar to the odd dimensional cases, only now for both $\psi_L$ and $\psi_R$ independently.  Hence, for $d = 2,6$ (mod $8$), when $\rho_L$ and $\rho_R$ are both pseudoreal the chiral symmetry group of $\mathcal{L}$ is given by $Sp(N) \times Sp(N)$, while when $\rho_L$ and $\rho_R$ are both real the chiral symmetry group is $O(2N) \times O(2N)$.  When $r$ is complex, the chiral symmetry group is $U(N) \times U(N)$.

For the cases in which $d = 4,8$ (mod $8$), both $s_L$ and $s_R$ are complex.  When $r$ is also complex, the reasoning is identical to that above for the $d = 2,6$ (mod $8$) cases with a complex representation of $G$ --- in such cases the chiral symmetry group is again given by $U(N) \times U(N)$.  When $r$ is either real or pseudoreal, denote by $\tilde{J}$ the charge conjugation operator restricted to the carrier space of $\rho_R$.  $\tilde{J}$~is then an antilinear map which intertwines $\rho_L$ with $\rho_R$ (i.e.~$\rho_L(g)\tilde{J} = \tilde{J} \rho_R(g)$ for all $g \in Spin(1,d-1) \times G$) and acts as complex conjugation on flavor space.  Using this, one can show that chiral symmetry transformations (in the Weyl basis) must all be of the form $\left( \begin{array}{cc}
X_1 &  Y_1 \tilde{J} \\
Y_2 \tilde{J}^{-1} & X_2
\end{array} \right)$,  
where $X_1$, $X_2$, $Y_1$, and $Y_2$ are (standard) linear operators which act non-trivially only on flavor space.  Such a transformation leaves $\mathcal{L}$ invariant if and only if
\begin{align}
\label{even48constraints}
X_1^{\dagger} X_1 + Y_2^{T} Y_2^{*} = I, \quad &  X_2^{\dagger} X_2 + Y_1^{T} Y_1^{*} = I  \\
 X_1^{\dagger} Y_1 \  + \   Y_2^{T} &  X_2^{*} = 0. \nonumber
\end{align}
One can then show that the set of transformations which satisfy the above constraints form a group isomorphic to $U(2N)$.

\section{Spontaneous Breaking of Chiral Symmetry}
\label{massterm}

\subsection{Assumptions}

Since we are considering confining gauge theories, we expect fermion condensates to be generated dynamically, signaling the spontaneous breaking of the chiral symmetries of $\mathcal{L}$.  That is, in any such gauge theory we expect some operators (constructed from fermion fields) which are not chirally invariant to acquire non-zero vacuum expectation values (VEVs).  Since we do not know the precise form of these operators, we make the following plausible assumptions ($\grave{a}$ la Peskin \cite{peskin_1980,peskin_slac}): 

\emph{1) Lorentz and Gauge Invariance}:  We assume that the vacuum is invariant under ${Spin(1,d-1) \times G}$.  This allows us, without loss of generality, to take any fermion condensate to transform trivially under ${Spin (1,d-1) \times G}$.  

\emph{2) Mass Terms}:  We assume that the residual symmetry of the theory after dynamical breaking can be determined from a \emph{single} non-chirally invariant Hermitian operator (not necessarily unique) which acquires a non-zero VEV, and which is a quadratic form in $\psi$ constructed from either a \emph{bilinear form} (linear in both arguments) or a \emph{Hermitian form} (linear in one argument and conjugate linear in the other) on the carrier space of~$\rho$.  (Moreover, we assume that such an operator depends on space-time coordinates only through~$\psi$.)   We hereafter refer to any such operator as a $\emph{mass term}$, since the non-zero VEV signals that (at least) some of the fermion flavors acquire a dynamically generated mass.

\emph{3) Flavor Democracy}:  We assume that \emph{all} of the fermion flavors acquire a mass; i.e.~any mass term originates from a nondegenerate form.  In $d = 4$, Coleman and Witten~\cite{witten_coleman} show that this is the case (given the above assumptions) in the large $N_c$ limit for $G = SU(N_c)$ when $r$ is the fundamental representation.  They also indicate that the argument is extendable to the gauge groups $SO(N_c)$ and $Sp(N_c)$.  We assume that their result further extends to any $d$, $G$ and $r$.  (For~a discussion of the $d=3$ case, see \cite{vafa_witten_1984b}.)  However, even given a situation in which not all fermions acquire mass, we could simply ignore the sector which remained massless;  the remaining fermion fields would then satisfy this assumption.
 
\emph{4) Parity Invariance}:  In $d = 4$, the Vafa-Witten theorem \cite{vafa_witten_1984,azcoiti_2008} shows that any mass term is parity invariant.  It has also been argued that the same is true for a complex representation of $G$ in $d=3$ \cite{appelquist_1995,appelquist_1990}.  We assume that any mass term is parity invariant for all $d$, $G$, and $r$; however, this is only necessary for our analysis when $d$ is odd.  

\vskip 4pt
In what follows, by a \emph{candidate operator} we will mean a Hermitian operator invariant under $Spin(1,d-1) \times G$ and parity which is a non-degenerate quadratic form in~$\psi$ (and depends on space-time coordinates only through $\psi$).  Clearly, by 1--4 above, any mass term is a candidate operator; however, we will need one further condition to determine those candidate operators which have the same residual symmetry as any mass term (i.e.~which have the residual symmetry of the gauge theory).
\vskip 4pt

\emph{5) Minimal Breaking}:  We assume that any mass term has the maximum residual symmetry among candidate operators.\footnote{One may worry that the set of residual symmetry groups of the candidate operators at a given $d$ and $r$ does not have a maximal element under inclusion, but this worry turns out to be unfounded.  (We have actually determined the residual symmetry groups for all candidate operators.  The results will appear in \cite{us}.)}   For $d=4$, this result was shown to hold for any $G$ and $r$ for fermions with a non-zero bare mass \cite{vafa_witten_1983}, and it is plausible that it continues to hold in the limit of massless fermions.   For the specific example of massless fermions carrying the fundamental representation of $G = SU(N_c), SO(N_c)$, or $Sp(N_c)$ in the large $N_c$ limit, Coleman and Witten \cite{witten_coleman} have demonstrated that the pattern of minimal breaking is indeed followed.  Their argument extends to all other even dimensional cases with $G = SU(N_c)$, as well as to the $d = 4,8$ (mod $8$) cases with $G = SO(N_c)$ or $Sp(N_c)$.  Although the techniques employed by Coleman and Witten do not extend to other cases, we expect that the pattern of minimal symmetry breaking is followed in complete generality.     

Before working out the consequences of these assumptions, we note that when we discuss the spontaneous breaking of chiral symmetries of $\mathcal{L}$, we necessarily exclude the $d = 1$ and $d = 2$ cases since the Coleman--Mermin--Wagner theorem \cite{coleman_1973} shows that there can be no spontaneous breaking of a continuous global symmetry for quantum field theories in $d \leq 2$ dimensions.  Thus, when we refer to $d = 1,2$ (mod $8$) in the context of symmetry breaking, we implicitly mean for the specific cases $d = 1,2$ to be ignored.  Moreover, for $d \geq 3$ one may expect there to be a critical number of flavors $N_0$ (which depends on $G$ and $r$) beyond which the theory will no longer be confining and no fermion condensate will form. (For a discussion of the well-known $d=4$ case, see~\cite{appelquist_1998}.  For $d=3$, see~\cite{appelquist_1995,appelquist_1990}.)  Since we only consider confining theories, we implicitly assume that $N \leq N_0$ in what follows.

\subsection{Candidate operators}

We first construct all the candidate operators; these depend on both $d$ and $r$.  In order to account for the $d$ dependence, we will apply Shaw's classification system \cite{shaw_1986} to the irreducible spinor representations.  For those unfamiliar, we give a brief account of Shaw's taxonomy.

For any finite-dimensional representation $\rho_0$ of a group $G_0$ on a complex vector space $\mathcal{V}$, we may naturally construct three additional representations.  These are the conjugate representation ($\bar{\rho}_0$), the contragredient (or transpose) representation ($\hat{\rho}_0$), and the contragredient of the conjugate representation ($\hat{\bar{\rho}}_0$).  Shaw's classification is based upon whether or not any of these four representations are equivalent, as well as the properties of certain antilinear operators and/or nondegenerate forms associated with such equivalences.  In the following, when two representations $\rho_1$ and $\rho_2$ are equivalent we will write $\rho_1 \simeq \rho_2$.

An irreducible representation $\rho_0$ is said to be class $0$ if none of the four representations above are equivalent to each other, in which case there is no invariant form or equivariant antilinear isomorphism on $\mathcal{V}$.  If both ${\rho_0 \simeq \hat{\rho}_0}$ and $\bar{\rho}_0 \simeq \hat{\bar{\rho}}_0$ (and these are the only equivalences), then $\rho$ is class~I.  In this case there is a unique (up to a scalar) invariant bilinear form $B$ on~$\mathcal{V}$.  If $B$ is symmetric (antisymmetric) then $\rho_0$ is in subclass I$_+$ (I$_-$).  Continuing to the next possibility, if the only equivalences are $\rho_0 \simeq \hat{\bar{\rho}}_0$ and ${\hat{\rho}_0 \simeq \bar{\rho}_0}$, then $\rho_0$ is class~II and there exists a unique (up to a scalar) invariant Hermitian form on $\mathcal{V}$.    The next case is when the only equivalences are $\rho_0 \simeq \bar{\rho}_0$ and $\hat{\rho}_0 \simeq \hat{\bar{\rho}}_0$; here $\rho_0$ is said to be class III, and there exists a unique (up to a phase) antilinear automorphism $J$ on $\mathcal{V}$ which commutes with $\rho_0(g)$ for all $g \in G_0$.  If $J^2 = I$ ($J^2 = -I$), then $\rho_0$ is in subclass III$_+$ (III$_-$).  The final possibility is when all four representations are equivalent, and hence all three structures exist (the Hermitian/bilinear form as well as the antilinear isomorphism) and are related.  In~this case $\rho_0$ is class IV, which divides into subclasses IV$_{\alpha \beta}$, where $\alpha, \beta~\in~\{ +, - \}$ and $\alpha$ is determined by $J$ and $\beta$ by $B$ in the obvious way.  Any irreducible $\rho_0$ is associated with exactly one of the Shaw classes discussed above. 

Shaw's system is a refinement of the classification of irreducible representations as real, pseudoreal, or complex.  Classes 0, I and II are complex, while III$_+$, IV$_{+ \beta}$ are real, and III$_{-}$, IV$_{-\beta}$ are pseudoreal.   When $G_0$ is a compact Lie group, one can say more.  In this case $\rho_0$ is equivalent to a unitary representation, and hence there is always an invariant Hermitian form on $\mathcal{V}$.  From this we see immediately that $\rho_0$ must be class II or IV, and with a little more work one can show that if $\rho$ is class IV, then it must be in the subclass IV$_{++}$ or IV$_{--}$.  Hence $\rho_0$ is class II if and only if it is complex, and $\rho_0$ is in the subclass IV$_{++}$ (IV$_{--}$) if and only if it is real (pseudoreal).  Since $Spin(1,d-1)$ is not compact, there are more possibilities for the Shaw classes of its representations.  We have computed the Shaw subclasses for all of the irreducible spinor representations, and the results are collected in Table \ref{ISRs}.

\begin{table}
\centering
\begin{tabular}{|c|c|c|c|}
\hline
$d$ mod 8 & type of IR & Shaw class & \# IR \\
\hline
$1$ & $\mathbb{R}$ & IV$_{++}$ & $1$ \\
\hline
$2$ & $\mathbb{R}$ & III$_{+}$ & $2$ \\
\hline
$3$ & $\mathbb{R}$ & IV$_{+-}$ & $1$ \\
\hline
$4$ & $\mathbb{C}$ & I$_{-}$ & $2$ \\
\hline
$5$ & $\mathbb{H}$ & IV$_{--}$ & $1$ \\
\hline
$6$ & $\mathbb{H}$ & III$_{-}$ & $2$ \\
\hline
$7$ & $\mathbb{H}$ & IV$_{-+}$ & $1$ \\
\hline
$8$ & $\mathbb{C}$ & I$_{+}$ & $2$ \\
\hline
\end{tabular}
\caption{Classification of the irreducible representations (IRs) of the group $Spin(1,d-1)$.  In the ``type of IR'' column, $\mathbb{R}$ denotes a real representation, $\mathbb{H}$ denotes a pseudoreal representation, and $\mathbb{C}$ denotes a complex representation.  The column ``\#IR'' gives the number of inequivalent IRs.
\label{ISRs}
}
\end{table}

We now return to the construction of the candidate operators.  In what follows, we explicitly discuss only the cases in which $d$ is odd and $r$ is either real or pseudoreal.  (The other cases can be treated similarly.)  In each of the odd dimensional cases we see that the Shaw class of $s$~is~IV.  As such, whenever $r$ is either real or pseudoreal, there exists both a unique\footnote{For clarity, we will treat the Hermitian/bilinear forms, as well as the antilinear maps, as unique in the following discussion, even though (as mentioned above) they are only unique up to scalar multiples.  None of our results will depend on the choice of these scalars.} nondegenerate invariant bilinear form and a unique nondegenerate invariant Hermitian form associated with $\rho$, as well as an invertible equivariant antilinear map $J$ on the carrier space of $\rho$ \cite{shaw_1986}.   

It is easy to see that $\gamma^{0}$ is associated with a nondegenerate Hermitian form on the spinor space, and since $s$ is irreducible, this form is unique.  Additionally, since $r$ is an irreducible unitary representation, the unique invariant Hermitian form on color space is simply the standard inner product.  Since $\rho$ is trivial on ``flavor'' space, the most general candidate operator which is a Hermitian form on the full representation space is given by $\psi^{\dagger} \gamma^{0} F \psi$, where $F$ is a linear operator which acts trivially on spinor and color spaces and is such that $F^{\dagger} = F$ and $F^{-1}$ exists, as required by the Hermiticity and non-degeneracy of candidate operators, respectively.  Also, recalling that $F$ is an operator on the ``flavor'' space $\mathbb{C}^N \otimes \mathbb{C}^2$, we note that the assumption of parity invariance requires that $\{F, I \otimes \sigma_1 \} = 0$.

We can construct an invariant bilinear form from the Hermitian form and the map $J$ above.  Namely, the most general candidate operator which is a bilinear form is given by $(J\psi)^{\dagger} \gamma^{0} F \psi + \textrm{h.c.}$, where $F$ acts non-trivially only on ``flavor'' space and can be taken to satisfy\footnote{For a quadratic form $\psi^T X \psi$ on Grassmann objects $\psi$, only the antisymmetric part of $X$ contributes.} $F = \pm F^T$, where the $\pm$ signs depend on both $d$ and $r$ as in Table \ref{pos_masses}.  (Similar to the Hermitian case, parity invariance requires that $\{F, I \otimes \sigma_1 \} = 0$, while non-degeneracy requires that $F$ be invertible.)

A similar analysis can be performed for all $d$ and $r$.  The results are summarized in Table \ref{pos_masses}.

\begin{table}
\centering
\footnotesize
{
\setlength{\extrarowheight}{2pt}
\begin{tabular}{|l|c|c|}
\hline
$d$ (type of $r$) & candidate operators  & $F, \gamma$ satisfies \\
\hline
\multirow{2}{*}{odd ($\mathbb{C}, \mathbb{R}, \mathbb{H})$}  & \multirow{2}{*}{$\psi^\dag \gamma^0 F \psi$} & $\{F, I \otimes \sigma_1 \} = 0$  \\
 &   & and $F = F^\dag$  \\
\hline
 1,7 \emph{mod} 8 ($\mathbb{R}$) & \multirow{2}{*}{$(J \psi)^\dag \gamma^0 F \psi + \textrm{h.c.}$}   & $\{F, I \otimes \sigma_1 \} = 0$  \\
or 3,5 \emph{mod} 8 ($\mathbb{H}$) &  & and $F^T = - F$  \\ 
\hline
 1,7 \emph{mod} 8 ($\mathbb{H}$) & \multirow{2}{*}{$(J \psi)^\dag \gamma^0 F \psi + \textrm{h.c.}$}   & $\{F, I \otimes \sigma_1 \} = 0$  \\
or 3,5 \emph{mod} 8 ($\mathbb{R}$) &  & and $F^T =  F$  \\
\hline
even ($\mathbb{C},\mathbb{R}, \mathbb{H})$ & $\psi_L^\dag \gamma F \psi_R + \textrm{h.c.}$ & $\scriptstyle \gamma^0 = \left( \begin{array}{cc} \scriptstyle 0 & \scriptstyle \gamma \\ \scriptstyle \gamma^\dag & \scriptstyle 0 \end{array} \right)$ \\
\hline
2,6 \emph{mod} 8 ($\mathbb{R}, \mathbb{H}$) & $(J_L \psi_L)^\dag \gamma F \psi_R + \textrm{h.c.}$ & $\scriptstyle \gamma^0 = \left( \begin{array}{cc} \scriptstyle 0 & \scriptstyle \gamma \\ \scriptstyle \gamma^\dag & \scriptstyle 0 \end{array} \right)$  \\
\hline 
\multirow{2}{*}{4,8 \emph{mod} 8 ($\mathbb{R}, \mathbb{H}$)} & \raisebox{2pt}{$\psi_L^\dag (\tilde{J} \gamma^\dag F_1 \psi_L + \gamma F_2 \psi_R)$} & \multirow{2}{*}{$\scriptstyle \gamma^0 = \left( \begin{array}{cc} \scriptstyle 0 & \scriptstyle \gamma \\ \scriptstyle \gamma^\dag & \scriptstyle 0 \end{array} \right)$} \\
 & \raisebox{2pt}{$+ \ \psi_R^\dag \tilde{J}^{-1} \gamma F_3 \psi_R + \  \textrm{h.c.}$} &  \\
\hline
\end{tabular}
}
\caption{All possible candidate operators for various $d$ and~$r$.  The~first column gives the space-time dimension $d$, and the type of the representation~$r$.  The second shows the form of an allowed candidate operator for that case, while the third shows any constraints that must be satisfied.  (Here $\gamma^{0}$ is shown in the Weyl basis.)  Note that in all cases $F$ (as well as $F_i$) acts non-trivially only on the appropriate flavor space, and $F^{-1}$ exists.  (For the final row, rewriting the mass term as $\psi^\dag Q \psi$, we have that $Q^{-1}$ exists.)
\label{pos_masses}
}
\end{table}

\subsection{Symmetry breaking patterns}

We now impose the assumption of minimal breaking to obtain the residual symmetries.  In what follows, we explicitly discuss only the cases in which $d = 3,7$ (mod $8$) and $r$ is either real or pseudoreal.  Here both bilinear and Hermitian forms exist, and one can show they give the same maximal residual symmetry; therefore we will restrict our discussion to the Hermitian form.  In this case, we have from Table \ref{pos_masses} that the most general candidate operator is given by $\psi^\dag \gamma^0 F \psi$.  A chiral symmetry transformation $X + Y J$ of $\mathcal{L}$ preserves this operator if and only if 
\begin{equation}
\label{oddrealprconstraintsmassterm}
X = FXF^{-1}  \quad \textrm{and} \quad  Y = FY(F^{*})^{-1}.  
\end{equation}
Using $\{ F ,I \otimes \sigma_1 \} = 0$, as well as the Hermiticity of $F$, one can show that for $X$ and $Y$ to satisfy the constraints in (\ref{oddrealprconstraintsmassterm}) they must be block diagonal.\footnote{Where we use the basis in which $I \otimes \sigma_1 = \left( \begin{array}{cc} 0 & I \\
I & 0 \end{array} \right)$.}  Moreover, one can easily see that for $F = I \otimes \sigma_3$ \emph{all} block diagonal $X$ and $Y$ satisfy (\ref{oddrealprconstraintsmassterm}), and so this $F$ will give the largest residual symmetry (and therefore the same residual symmetry as any mass term).  Hence, when $\rho$ is real, the residual symmetry group is isomorphic to $O(2N) \times O(2N)$; similarly, when $\rho$ is pseudoreal, we obtain $Sp(N) \times Sp(N)$.  Thus, in the former case the symmetry breaking pattern is $O(4N) \longrightarrow O(2N) \times O(2N)$, while in the latter case we have $Sp(2N) \longrightarrow Sp(N) \times Sp(N)$.

A similar analysis can be performed for all $d$ and $r$, and the results are summarized in Table \ref{residualsymmetrytable}.  The candidate operators which yield the maximal residual symmetry are not unique; below we provide one choice for each $d$ and~$r$ (using the notation of Table \ref{pos_masses}).  For any odd $d$, the maximal residual symmetry is obtained for the Hermitian form with $F = I \otimes \sigma_3$. For $d = 2,6$ mod 8 and any type of representation of $G$, as well as $d = 4,8$ mod 8 and a complex representation of $G$, the maximal residual symmetry is obtained for the Hermitian form with $F = I$.
For the cases $d = 4$~mod~8 with $r$ pseudoreal and $d = 8$ mod 8 with $r$ real, taking $F_1 = F_3 = I$ and $F_3 = 0$ yields the maximal residual symmetry.  Finally, when $d = 4$ mod 8 with $r$ real or when $d = 8$ mod~8 with $r$ pseudoreal, the maximal residual symmetry is obtained for $F_1 = F_3 = 0$ and $F_2 = iI$.

\begin{table}
\centering
\footnotesize
\begin{tabular}{|c|c|c|c|}
\hline
$d$ & IR of $G = \mathbb{C}$ & IR of $G = \mathbb{R}$  & IR of $G = \mathbb{H}$ \\
\hline
$1$ & $U(2N) \longrightarrow$ & $O(4N) \longrightarrow$ & $Sp(2N) \longrightarrow$ \\
 & $U(N) \times U(N)$ & $U(2N)$ & $U(2N)$ \\
\hline
$2$ & $U(N) \times U(N)$ & $O(2N) \times O(2N)$ & $Sp(N) \times Sp(N)$ \\
 & $\longrightarrow U(N)$ & $\longrightarrow O(2N)$ & $\longrightarrow Sp(N)$ \\
\hline
$3$ & $U(2N) \longrightarrow$ & $O(4N) \longrightarrow$ & $Sp(2N) \longrightarrow$ \\
 & $U(N) \times U(N)$ & $O(2N) \times O(2N)$ & $Sp(N) \times Sp(N)$ \\
\hline
$4$ & $U(N) \times U(N)$ & $U(2N)$ & $U(2N)$ \\
 & $\longrightarrow U(N)$ & $\longrightarrow O(2N)$ & $\longrightarrow Sp(N)$ \\ 
\hline
$5$ & $U(2N) \longrightarrow$ & $Sp(2N) \longrightarrow$ & $O(4N) \longrightarrow$ \\
 & $U(N) \times U(N)$ & $U(2N)$ & $U(2N)$ \\
\hline
$6$ & $U(N) \times U(N)$ & $Sp(N) \times Sp(N)$ & $O(2N) \times O(2N)$ \\
 & $\longrightarrow U(N)$ & $\longrightarrow Sp(N)$ & $\longrightarrow O(2N)$ \\
\hline
$7$ & $U(2N) \longrightarrow$ & $Sp(2N) \longrightarrow$ & $O(4N) \longrightarrow$ \\
 & $U(N) \times U(N)$ & $Sp(N) \times Sp(N)$ & $O(2N) \times O(2N)$ \\
\hline
$8$ & $U(N) \times U(N)$ & $U(2N)$ & $U(2N)$ \\
 & $\longrightarrow U(N)$ & $\longrightarrow Sp(N)$ & $\longrightarrow O(2N)$ \\
\hline
\end{tabular}
\caption{The chiral symmetry group of $\mathcal{L}$ and the residual symmetry for each space-time dimension $d$ (mod 8) in the cases in which the irreducible representation (IR) of the gauge group $G$ carried by the fermions is complex, real, and pseudoreal.  
\label{residualsymmetrytable}
}
\end{table}

\section{Bott periodicity and the low-energy effective theory}

\subsection{Configuration space topology}

Now that we know the pattern of chiral symmetry breaking in our gauge theories, we can investigate the low-energy effective theory for the associated Goldstone boson degrees of freedom. This effective theory is a nonlinear sigma model whose fields map the space-time $\mathbb{R}^d$ into the coset space ${\hat G}/H$, where ${\hat G}$ is the chiral symmetry group of ${\cal L}$, and $H$ the residual symmetry group after dynamical breaking. The {\it classical configuration space} of the model is the set of all finite-energy fixed-time fields, each of which can be viewed as a (continuous) map from the $(d-1)$-sphere $S^{d-1}$ to ${\hat G}/H$, where the spatial manifold $\mathbb{R}^{d-1}$ has been one-point compactified to $S^{d-1}$ by the finite energy condition. We also assume that the ``point at infinity" on $S^{d-1}$ always maps to a fixed point on ${\hat G}/H$, independent of the field configuration.\footnote{Any path between two field configurations which map the point at infinity to distinct points in ${\hat G}/H$ will have infinite action.} (The constant map from $S^{d-1}$ to this point represents the unique vacuum configuration.)  Let us denote this configuration space by ${\rm Map}_*(S^{d-1},{\hat G}/H)$, which we endow with the compact-open topology.\footnote{Restricting ourselves to the subspace of differentiable (or even smooth) maps in ${\rm Map}_*$ would not change any of our results below.} 

Denote by $M^{N,\lambda}_d$ (with $\lambda=\mathbb{R} ,\mathbb{C}$ or $\mathbb{H}$) the above coset space for our $d$-dimensional gauge theory with $N$ flavors of massless fermions in a representation of the gauge group of type $\lambda$. (For example, $M^{N,\mathbb{R}}_5=Sp(2N)/U(2N)$.) From Table 3, we clearly have that 
$$M^{N,\mathbb{C}}_d=M^{N,\mathbb{C}}_{d+2}$$
$$M^{N,\lambda}_d=M^{N,\lambda}_{d+8} \quad \textrm{ for }  \lambda=\mathbb{R},\mathbb{H}$$
$$M^{N,\mathbb{R}}_d=M^{N,\mathbb{H}}_{d+4} . $$
In this context, the \emph{Bott periodicity theorem} \cite{Bott} can be taken as a statement relating the homotopy type\footnote{Recall that two topological spaces $X$ and $Y$ are said to be {\it homotopically equivalent} (or to have the same {\it homotopy type}) if there exist continuous maps $f:X\to Y$ and $g:Y\to X$ such that the composites $g\circ f$ and $f \circ g$ are homotopic to the identity maps on $X$ and $Y$, respectively. In particular, the map $f$ (or $g$) induces isomorphisms between the homotopy groups of $X$ and those of $Y$ in any dimension. The (co)homology groups of $X$ and $Y$ are similarly isomorphic.} of these coset spaces as ${N\to\infty}$. More specifically, 
$$\Omega M^{\infty ,\mathbb{C}}_{2d+1}\simeq M^{\infty ,\mathbb{C}}_{2d},\ \ \ \Omega M^{\infty ,\mathbb{C}}_{2d}\simeq \mathbb{Z}\times M^{\infty ,\mathbb{C}}_{2d-1}$$
$$\Omega M^{\infty ,\mathbb{R}}_d\simeq M^{\infty ,\mathbb{R}}_{d-1}\ \ (d\neq 0,4\ {\rm mod}\ 8)$$
$$\Omega M^{\infty ,\mathbb{R}}_d\simeq \mathbb{Z}\times M^{\infty ,\mathbb{R}}_{d-1}\ \ (d=0,4\ {\rm mod}\ 8)$$
where $\Omega X\equiv {\rm Map}_*(S^1,X)$ is the based loop space of the topological space $X$, ``$\simeq$" denotes homotopy equivalence, and $\mathbb{Z}$ represents the integers.

Now the configuration space of the $d$-dimensional gauge theory associated with $M^{N,\lambda}_d\,$ is $\,Q^{N,\lambda}_d={\rm Map}_*(S^{d-1},M^{N,\lambda}_d)$ $\simeq \Omega^{d-1}M^{N,\lambda}_d$ (the ($d-1$)-fold iterated loop space). The statement of the periodicity theorem above then gives, for {\it any} $d$, that 
$$Q^{\infty,\mathbb{C}}_d\simeq \mathbb{Z} \times M^{\infty,\mathbb{C}}_1=\mathbb{Z}\times (U/(U\times U))\simeq \Omega U$$
$$Q^{\infty,\mathbb{R}}_d\simeq M^{\infty,\mathbb{R}}_1=O/U \simeq \Omega O$$
$$Q^{\infty,\mathbb{H}}_d\simeq M^{\infty,\mathbb{H}}_1=Sp/U \simeq \Omega Sp$$
where $O$, $U$, and $SP$ are the direct limits (as $n\to\infty$) of $O(n)$, $U(n)$ and $Sp(n)$ respectively.  So even though the pattern of chiral symmetries and their breaking varies as $d$ changes, Bott periodicity shows us that, for fixed $\lambda$, the homotopy type of the configuration space of the low-energy effective theory is independent of $d$ as the number of massless fermion flavors tends to infinity.

\subsection{Baryons as topological solitons}

It is well-known that the low-dimensional homotopy and (co)homology groups of $Q^{N,\lambda}_d$ have direct physical relevance. For example, $\pi_0(Q^{N,\lambda}_d)$ counts the distinct path-components of $Q^{N,\lambda}_d$. Hence, if $\pi_0(Q^{N,\lambda}_d)$ is non-trivial, then there are fields that cannot be continuously deformed into the vacuum configuration (the constant field). These different path-components represent superselection sectors in the associated  quantum theories, and their existence suggests that the model may possess topological solitons --- that is, particle-like solutions of the classical equations of motion which cannot be deformed into the vacuum. (Whether or not such solutions exist, and whether they are dynamically stable, will depend on the details of the effective Lagrangian for the sigma model.) Upon quantization, these classical solitons give rise to particle-like quantum states with a conserved ``topological charge" labeled by the elements of $\pi_0(Q^{N,\lambda}_d)$. 

From the above characterization of the homotopy type of $Q^{\infty,\lambda}_d$ (and standard results in algebraic topology) we have $\pi_0(Q^{\infty,\mathbb{C}}_d)=\pi_1(U)=\mathbb{Z}$, $\pi_0(Q^{\infty,\mathbb{R}}_d)=\pi_1(O)=\mathbb{Z}_2$, and $\pi_0(Q^{\infty,\mathbb{H}}_d)=\pi_1(Sp) = \{e\}$ (the trivial group).
This is consistent with the expectation that baryons in our gauge theories should show up in the low-energy effective theory as topological solitons (at least at large enough $N$), with the associated baryon number being the conserved topological charge (as happens in $d=4$) \cite{witt}. For concreteness, let us choose our gauge groups to be $SU(N_c)$, $SO(N_c)$, and $Sp(N_c)$ for $\lambda =\mathbb{C},\mathbb{R}$, and $\mathbb{H}$, respectively (with $N_c \geq 3$ for $\lambda = \mathbb{C},\mathbb{R}$), and let each fermion flavor transform as the fundamental representation (we will call these fermions ``quarks" in what follows). For $\lambda =\mathbb{C}$ and~$\mathbb{R}$, color singlet ``baryons" can be made from $N_c$ quarks, and ``anti-baryons'' from $N_c$ anti-quarks. For $\lambda=\mathbb{C}$ these two possibilities are distinct and stable, corresponding to baryon number ${B=+1}$ and $B=-1$, respectively. This allows us to construct states with any integer baryon number $B$, and $B$ will be conserved. (Compare to $\pi_0(Q^{\infty,\mathbb{C}}_d)=\mathbb{Z}$.) For $\lambda=\mathbb{R}$ there is no distinction between quark and anti-quark (since here $r \simeq \bar{r}$), and hence no distinction between baryon and anti-baryon. Any $B=2$ state can now decay into $N_c$ mesons, so that baryon number will only be conserved mod 2~\cite{witt}. ($\pi_0(Q^{\infty,\mathbb{R}}_d)=\mathbb{Z}_2$.) Finally, for $\lambda=\mathbb{H}$ there is again no distinction between quark and anti-quark, but now there are no baryons at all since any candidate $B=1$ state (which here contains $2N_c $ quarks) can decay into $N_c$ mesons. Thus, there is no conserved baryonic quantum number~\cite{witt}. ($\pi_0(Q^{\infty,\mathbb{H}}_d)=\{e\}$.) 

Proceeding to higher homotopy groups of $Q^{\infty,\lambda}_d$, it is straightforward to show that each path component of $Q^{\infty,\lambda}_d$ is simply-connected for any $\lambda$; that is, $\pi_1(Q^{\infty,\lambda}_d)=\{e\}$. Hence, there are no possible ``$\theta$-vacua" in these sigma models. (More precisely, there are no non-trivial flat vector bundles over $Q^{\infty,\lambda}_d$.) Finally, in any path component of $Q^{\infty,\lambda}_d$ we have that  $\pi_2(Q^{\infty,\lambda}_d)=H^2(Q^{\infty,\lambda}_d;\mathbb{Z})=\mathbb{Z}$ for any $\lambda$, showing the formal existence of certain topological terms, with quantized coefficients, available for our sigma models. (Equivalently, there exist complex line bundles over $Q^{\infty,\lambda}_d$ with first Chern class of infinite order.) For $d$ even, these are the standard Wess-Zumino terms for the ${\hat G}/H$ sigma models, while for $d$ odd, they are nonlinear realizations of the Chern-Simons term for the ``hidden local symmetry group" $H$ in the ${\hat G}/H$ sigma model. For the concrete gauge groups and fermion representations in the preceding paragraph (and any $d$), we expect that these terms will be generated in the low-energy effective theory with a coefficient proportional to $N_c$, which for $\lambda =\mathbb{C}$ and~$\mathbb{R}$ would lead to the solitons with unit topological charge (that is, baryon number) being (spinorial) fermions for $N_c$ odd, and (tensorial) bosons for $N_c$ even, in a manner similar to the well-known situation in $d=4$ \cite{witt}. More precisely, when evaluated on a time-dependent configuration which rotates a $B=1$ soliton by $2\pi$ (or exchanges two identical $B=1$ solitons), this term evaluates to $N_c\,\pi$, and hence makes a contribution to the path integral of $(-1)^{N_c}$. We will provide a more detailed analysis of the baryon number, spin, and statistics of these topological solitons in \cite{us}.\footnote{A similar analysis has already been performed for (2+1)-dimensional QCD in \cite{rajeev,imbo_1991}.}

We close this section with two comments. 

\vskip 4pt
\noindent
(1) One may worry that by letting $N\to\infty$ in the above analysis we have entered the non-confining regime. However, many of the computations (and physical interpretations) of the low-dimensional homotopy groups of $Q^{N,\lambda}_d$ given above did not require $N\to\infty$. Indeed, they often hold starting at relatively small values of $N$. Using $N\to\infty$ simply allowed us to state the results more compactly and elegantly. Moreover, for the specific examples with $G=SU(N_c)$ and $SO(N_c)$ considered above, the baryons behave precisely as topological solitons only in the limit $N_c\to\infty$ \cite{witten_1979}. In this limit, we expect the critical number of flavors $N_0$  to be proportional to $N_c$. (For a discussion in $2+1$ dimensions, see \cite{appelquist_1990}.  For $3+1$ dimensions see \cite{appelquist_1998}.) Hence, letting $N$ be large in this case may still be safe for confinement and chiral symmetry breaking. 

\vskip 4pt
\noindent
(2) One may wonder if an analysis of the possible anomalous breaking of portions of ${\hat G}$ (which we ignore here) may ruin the nice picture that has emerged of the topological properties of the configuration space of the low-energy effective theory. However, we do not expect this to be the case. More specifically, we suspect that things work similarly to $(3+1)$-dimensional QCD where the anomalous breaking of the axial $U(1)$ subgroup of ${\hat G}=U(N)\times U(N)$ does not lead to any change in the homotopy type of the configuration space of the associated sigma model. 

\vskip 4pt
\noindent
We hope to say more about both of these issues in~\cite{us}.

\section{Conclusions}

We have computed the group of chiral symmetries of the Lagrangian for confining vector-like gauge theories with massless fermions in $d$-dimensional Minkowski space and, under a few simple assumptions, determined the form of the relevant fermion condensates, as well as the residual symmetries after spontaneous breaking.  These realizations of chiral symmetry follow the pattern of Bott periodicity across dimensions.  When the fermions carry a complex representation of the gauge group, this pattern has periodicity~$2$, while for a real or pseudoreal representation of the gauge group, the pattern has periodicity~$8$.  Moreover, the patterns for the real and pseudoreal cases are shifted by $4$ relative to one another.  The chiral symmetries and their breaking are recounted in Table \ref{residualsymmetrytable}. It then follows from the Bott periodicity theorem that the homotopy type of the configuration space of the low energy effective theory (at fixed gauge group representation type) is independent of~$d$ in the large flavor limit. Simple computations then support the interpretation of baryons as topological solitons at low energies (in those cases where there is a conserved baryon number). This extends well-known results in $d=4$ to any space-time dimension.

In conclusion, our results further exemplify the deep connection between Bott periodicity and physical models involving spinors across dimensions.  This connection is well-known in string theory, and has also become apparent in the classification of topological insulators \cite{kitaev_2009,stone_2010}.  Chiral symmetry breaking in vector-like gauge theories is yet another area where we can see this deep result of pure mathematics influencing the form of physical theory.

\vskip 6pt
\noindent
{\it Acknowledgements:} We thank Mark Mueller for many insightful comments and contributions. We also thank Randall Espinoza, Ben Grinstein, Wai-Yee Keung, Ira Rothstein, Pedro Schwaller, Misha Stephanov, and Ho-Ung Yee for interesting discussions.


\end{document}